\documentclass[proceedings, preprint]{rmaa}

\usepackage{paralist}

\usepackage{psfrag,color}

\usepackage{amsmath}

\usepackage{hyperref}
\hypersetup{colorlinks=true,linkcolor=blue,citecolor=blue,filecolor=blue,urlcolor=blue}

\usepackage{tikz,xcolor,graphicx}
\graphicspath{{images/}}

\usepackage[normalem]{ulem}

\definecolor{lime}{HTML}{A6CE39}
\DeclareRobustCommand{\orcidicon}{%
    \begin{tikzpicture}
    \draw[lime, fill=lime] (0,0) 
    circle [radius=0.16] 
    node[white] {{\fontfamily{qag}\selectfont \tiny ID}};
    \draw[white, fill=white] (-0.0625,0.095) 
    circle [radius=0.007];
    \end{tikzpicture}
    \hspace{-2mm}
}

\newcommand{\orcidVP}{\href{https://orcid.org/0000-0002-3031-062X}{\orcidicon}}
\newcommand{\orcidTG}{\href{https://orcid.org/0000-0002-0592-489X}{\orcidicon}}
\newcommand{\orcidJV}{\href{https://orcid.org/0000-0002-1690-6072}{\orcidicon}}
\newcommand{\orcidRK}{\href{https://orcid.org/0009-0009-8595-5562}{\orcidicon}}

\SetYear{2024}

\SetConfTitle{III Workshop on Astronomy Beyond the Common Senses for Accessibility and Inclusion}

\title{Fostering innovation, inclusion, and diversity in astronomy education: The Czech Astronomy Olympiad experience} 

\author{
  V\'{a}clav Pavl\'{i}k \altaffilmark{1,2,3} \orcidVP,
  Jakub Vo\v{s}mera \altaffilmark{1,4} \orcidJV,
  Tom\'{a}\v{s} Gr\'{a}f \altaffilmark{1,5} \orcidTG,
  and Radka K\v{r}\'{i}\v{z}ov\'{a} \altaffilmark{1,6} \orcidRK
}

\altaffiltext{1}{Czech Astronomy Olympiad,
Czech Astronomical Society,
Fri\v{c}ova 298,
251~65, Ond\v{r}ejov, Czech Republic.\\
(First author's email: \href{mailto:pavlik@astro.cz}{pavlik@astro.cz})}

\altaffiltext{2}{Astronomical Institute of the Czech Academy of Sciences,
Bo\v{c}n\'i~II~1401,
141~31, Prague~4, Czech Republic.}

\altaffiltext{3}{Department of Astronomy,
Indiana University,
Swain Hall West,
727 E 3$^\text{rd}$ Street,
Bloomington, IN 47405, USA.}

\altaffiltext{4}{Institut de Physique Th\'{e}orique, 
CEA Paris-Saclay, CNRS, 
Orme des Merisiers, 
91191 Gif-sur-Yvette, France.}

\altaffiltext{5}{Silesian University in Opava,
Institute of Physics in Opava,
Bezru\v{c}ovo n\'{a}m\v{e}st\'{i} 13,
746~01, Opava, Czech Republic.}
 
\altaffiltext{6}{
Faculty of Mathematics and Physics, Charles University,
V Hole\v{s}ovi\v{c}k\'{a}ch 2, 180 00, Prague 8, Czech Republic.}

\shortauthor{Pavl\'{i}k et al.}
\shorttitle{Czech Astronomy Olympiad}

\listofauthors{V. Pavl\'{i}k, J. Vo\v{s}mera, T. Gr\'{a}f, \& R. K\v{r}\'{i}\v{z}ov\'{a}}
\indexauthor{Pavl\'{i}k, V.}
\indexauthor{Vo\v{s}mera, J.}
\indexauthor{Gr\'{a}f, T.}
\indexauthor{K\v{r}\'{i}\v{z}ov\'{a}, R.}

\abstract{%
Astronomy education and outreach are very important when it comes to the future generation's interest in science. The Czech Astronomy Olympiad shows how an educational competition for secondary and high schools can help us drive innovation and promote inclusion and diversity. In this work, we introduce the scope of this competition and show statistics on participation. We also discuss some of the steps taken to make astronomy accessible to a wider audience, such as organising international workshops. In addition, we explore some of the approaches which were adopted to broaden the Olympiad's reach and impact. These include, e.g., developing a dedicated website environment or publishing Open Access booklets with solved problems.
}

\resumen{%
La educación y la divulgación de la astronomía son muy importantes para que las futuras generaciones se interesen por la ciencia. La Olimpiada Checa de Astronomía muestra cómo una competición educativa para centros de secundaria y bachillerato puede ayudarnos a impulsar la innovación y promover la inclusión y la diversidad. En este trabajo, presentamos el alcance de esta competición y mostramos estadísticas sobre la participación. También mostramos algunas de las medidas que tomamos para hacer la astronomía más accesible a todos, como la organización de talleres internacionales. Por último, exploramos algunos de los enfoques que adoptamos para ampliar nuestro alcance e impacto, como el desarrollo de un entorno web dedicado o la publicación de folletos de acceso abierto con problemas resueltos.
}

\addkeyword{education: science education, inclusive education}
\addkeyword{online learning}
\addkeyword{sociology of astronomy}

\begin{document}
\maketitle

\section{Introduction}
\label{sec:introduction}

The Czech Astronomy Olympiad (hereafter CzAO) is an extra-curricular competition for secondary-school and high-school students. It was founded in 2003 and is currently in its twenty-first year. It is organised by the Czech Astronomical Society and is announced annually among the top-tier competitions by the Czech Ministry of Education, Youth and Sports. CzAO is traditionally part of the community of national competitions participating in the International Olympiad on Astronomy and Astrophysics (IOAA).

\section{Structure of the Olympiad}
\label{sec:structure}

\subsection{Czech Olympiad}

CzAO has four age categories: AB (for the oldest students), CD, EF, and GH (for the youngest ones) which is a similar classification to, e.g., the Physics Olympiad (with categories A--H). However, since astronomy is not part of the standard school curriculum in the Czech Republic (which makes the assignment of topics to individual school grades rather challenging), we are merging two consecutive school grades (e.g., AB for the two most senior ones). Consequently, students are allowed to compete in their age category for two years in a row which gives them further motivation to improve. All students are also allowed to attempt any higher-grade category which encourages the more advanced ones to showcase their abilities (it also sometimes happens that a younger student wins one or more of the older categories).

Over the course of one academic year, CzAO is divided into several rounds. It all begins with the \textit{school round} where the students solve multiple-choice questions and one or two computational problems. This round takes place in person at schools, is supervised by teachers and is constrained by a time limit. Its purpose is to introduce the students to astronomy, therefore, the exercises are easier and the students are allowed to use any resources (except for being helped by someone else), including books, notes, the internet, etc.

Those who succeed in the school round progress to the \textit{regional round} where they have up to three months to solve multiple-choice questions, several longer and more difficult computational problems, and, finally, an observational exercise. The students work on the regional round at home and send their solutions to the Board of CzAO for grading.

Approximately 25 top-ranked students from each category are then selected to compete in the country-wide \textit{final round}, organised in Prague (for categories EF, GH) and Opava (for AB, CD). Here, the students are faced with several multiple-choice questions and complicated computational problems. The high-school categories also include questions on data analysis and night-sky observations. All competitors also have to demonstrate their orientation abilities under the artificial sky in a planetarium.

\subsection{International Olympiads}

CzAO is fully integrated within the framework of the International Olympiad on Astronomy and Astrophysics (hereafter IOAA), a renowned student competition with worldwide recognition and nearly 20 years of tradition. We submit proposals for travel funding to the Ministry of Education, Youth and Sports annually to ensure that our best students from the categories AB, CD and EF can also participate in IOAA (primarily intended for 16 to 20-year-old students) or IOAA-junior (students younger than 16). Before IOAA had a junior category, our younger students would also participate in the International Astronomy Olympiad (IAO). See also Tab.~\ref{tab:intl}.

The selection process for these two international competitions usually takes place at the International Workshop on Astronomy and Astrophysics (IWAA), organised by CzAO and our international partners since 2016 \citep[see][]{iwaa_ccf}.\!\footnote{The workshops of CzAO are being financially supported by the Czech Ministry of Education, Youth and Sports within the ``Giftedness Strategy'' programme.} It is a week-long event where more than 40 students (not only from the Czech Republic but also from other nearby European countries) come together to participate in problem-solving-oriented training sessions, competition simulation and other non-astronomical and social activities. Through this effort, CzAO strives to foster collaboration and to exchange know-how between astronomy educators around the continent. Moreover, this also creates a unique environment for students to interact with their peers from other countries. Hence, they can share experience and culture, and learn that education and science transcend geographical borders.

\begin{table}[!t]
    \centering
    \small
    \setlength{\tabnotewidth}{\columnwidth}
    \tablecols{4}
    \setlength{\tabcolsep}{1\tabcolsep}
    \caption{International ranking\tabnotemark{a}}
    \begin{tabular}{cccc}
        \toprule
        Year & IAO   & IOAA  & IOAA-jr \\
        \midrule
        2007 & 1S, 4H & & \\
        2008 & 2B, 4H & & \\
        2009 & 2S, 3B & & \\
        2010 & & 1G & \\
        2011 & 1G, 1B, 3H & 1G\tabnotemark{b}, 1S, 1H & \\
        2012 & 1S, 4B, 1H & 2G, 2S, 1B & \\
        2013 & 3B, 2H & 1S, 1B, 2H & \\
        2014 & 1S, 1B, 3H & 2B, 2H & \\
        2015 & 2S, 1B, 2H & 3H & \\
        2016 & 1S, 2B, 3H & 1G, 1B, 3H & \\
        2017 & 1S, 1B, 3H & 1S, 2B, 1H & \\
        2018 & 1S, 3B, 1H & 1G, 1S, 3B & \\
        2019 & 3B, 2H & 1G, 1S, 2B & \\
        2020 &  & 2S, 4B, 2H\tabnotemark{c} & \\
        2021 & 2S, 3B, 1H & 3S, 7B & \\
        2022 & 1S, 5B, 1H & 3S, 1B, 1H & 2G, 1B \\
        2023 &  & 1G, 1S, 3B & 2G, 2S, 1B \\
        \bottomrule
        \tabnotetext{a}{Each field shows the number of gold (\textbf{G}), silver (\textbf{S}) and bronze (\textbf{B}) medals, and honourable mentions (\textbf{H}) earned by Czech students. The field is left empty in the cases when we did not participate.}
        \tabnotetext{b}{Our participant was the overall winner.}
        \tabnotetext{c}{IOAA was cancelled due to the pandemic, however, the IOAA organisers prepared the Global e-Competition on Astronomy and Astrophysics (GeCAA) instead.}
    \end{tabular}
    \label{tab:intl}
\end{table}

\section{Students' participation}

\begin{figure}[!t]
    \centering
    \includegraphics[width=\linewidth]{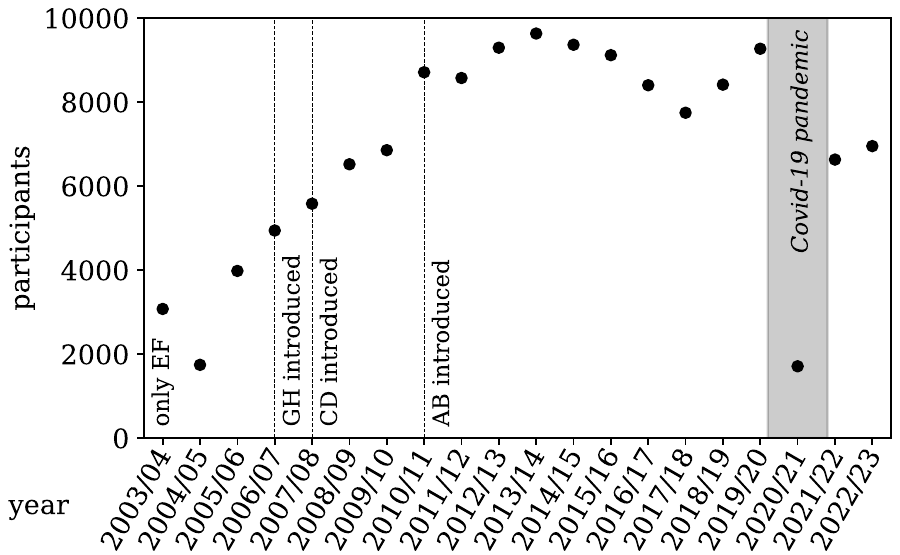}\\
    \vspace*{-10pt}
    \caption{The total number of participating students in the school round of CzAO through its history. The vertical lines show when each age category was introduced. The grey area shows the approximate span of the lockdown in our country due to the Covid-19 pandemic.}
    \label{fig:school}
\end{figure}
\begin{figure}[!t]
    \centering
    \includegraphics[width=\linewidth]{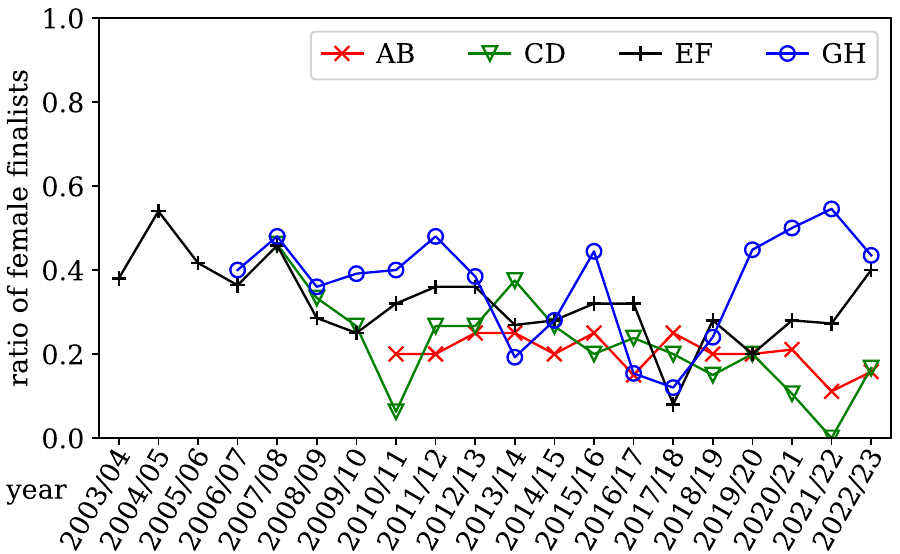}\\
    \vspace*{-10pt}
    \caption{The ratio of female participants in the national finals of CzAO in different categories.}
    \label{fig:final_women}
\end{figure}

Although astronomy is not taught as a subject in Czech secondary and high schools, CzAO is able to reach thousands of students every year and inspire many of them to follow astronomy and science as their interests or career paths. In Fig.~\ref{fig:school}, we show the overall participation since the beginning of CzAO. The EF category is historically the oldest, and other ones were gradually introduced later, as we also show in this figure. After the introduction of the last age category in 2010, we had a stable base of 8--10k students in the school round annually for the whole decade. Thereafter, we can see the negative effect of the Covid-19 pandemic between the years 2020 and 2021 where the overall participation dropped to about 18\,\% of the previous year. This was a global trend throughout all competitions in the Czech Republic because of the closure of schools and the limitation of funding. Nonetheless, we did not give up and organised the whole year online. Now we are steadily growing back to above 7k students per year.

We are also delighted to see a significant representation of women among our student competitors. For several years now, we have had around 40 to 50\,\% female students in the school rounds. In the last two years, it was even 54\,\% and 55\,\%, respectively, which shows that STEM education in our country is balanced at schools. The average number of women who progress to the national finals is then lower, between 20 to 40\,\% (see Fig.~\ref{fig:final_women}). Despite the large fluctuations between the years, it is apparent that younger categories usually have a higher percentage of female finalists than high school categories. This also propagates to the overall ranking of girls in the finals
where, on average, females achieve the best performance in the youngest category. Therefore, our current long-term goal is to promote astronomy among female high-school students.


\section{Team of organisers}

Traditionally, most organisers are recruited from former participants when they graduate from high school and go to university. This ensures continuous renewal of the team, diversity and professional quality of the competition problems. The organisers are also an inspiration to the participants, motivating them to continue to pursue STEM fields and, in many cases, enter scientific careers.

The representation of women in CzAO mirrors, to an extent, the national long-term average of women studying natural sciences, engineering or maths at Czech universities and pursuing careers in related research fields \citep[based on the report from][]{soc_avcr}. Nevertheless, although the majority of the CzAO organisers are men at the moment, we see an increase in the number of our female colleagues. We hope this upward trend will continue as our current female participants get older and will join us.

Our goal is to make CzAO accessible to everyone. Thus, we regularly submit proposals for dedicated funding from the Ministry of Education, Youth, and Sports, which allows us to organise all activities without imposing any financial burden on our students. Alongside the Olympiad itself, it enables us to host the IWAA or other annual workshops, go to international competitions and provide further mentorship to our top students. This helps them prepare for their future academic endeavours.

\section{Technological solutions}

\subsection{Website environment}

Throughout its existence, CzAO has been using online platforms\footnote{\url{https://olympiada.astro.cz/}} to inform students and teachers about the current and past years of the competition. In the old days, the teachers would log in, download and print the school-round exercises, distribute them among students, grade and submit the results. It also used to be the teachers who were administering the regional round exercises, collecting the students' solutions and mailing them to the Board of CzAO.

As our student base expanded, it became increasingly more difficult to grade and keep track of all the solutions in both the school and the regional rounds. 
In conjunction with the growing demand by schools to reduce the volume of printed materials, this motivated us to further develop our online capabilities. Since 2010, we have enabled the students to solve the school round on a computer \citep[see also][]{ict}. To make the regional round accessible to more participants, we also added a module which allowed them to solve the multiple-choice questions and enter the observational part online.

With the advent of the Covid-19 pandemic, the government suspended funding for all student competitions. The mandated lockdown
led to all schools being closed during the ongoing regional round. Moreover, the Board of CzAO could not meet to grade the regional round as usual. Nonetheless, our sophisticated IT system gave us an advantage in this situation -- we added another module for submitting scanned documents to our websites and enabled each student to upload their work online. The Board was then able to evaluate these scans remotely from home (not only from the Czech Republic but also from Germany, Switzerland and the USA). 

Overall, these initiatives greatly eased the processes of administering and evaluating the students' solutions. Thanks to our early transition to the online world, we were able to engage students from diverse geographical locations and eliminate the usual barriers associated with physical attendance in various phases of the competition.

\subsection{Typesetting}

CzAO originally used Microsoft Word to create documents with problems and model answers. While it served its purpose for a long time, as the organisational team grew, so did the variety of operating systems. We often could not achieve satisfactory results without licensing this proprietary editor. Therefore, since the academic year 2014/15, the Board of CzAO has decided to switch to \LaTeX\ -- a multi-platform, open-source language that is widely used for typesetting, e.g., research works. We have developed a custom style \texttt{astroolymp.sty}\!\footnote{Freely available at \url{https://github.com/pavlikva/astroolymp} under the GNU General Public License v3.0.} which allows us to produce all competition-related documents with ease.

\section{Publication activities}

The primary instruction language of CzAO is Czech. However, we are committed to making astronomical knowledge accessible to the global audience. Therefore, alongside the Czech media and conferences, we also publicise CzAO at international educational conferences \citep[e.g.,][]{ips2018}. Moreover, several years ago, we also started to translate our competition problems into English and publish them in open-access booklets 
\citep{ao2016,ao2017,ao2018,ao2019,ao2021,ao2022,ao2023}. This initiative was met with positive feedback from our international colleagues. Some foreign Olympiads even use them to prepare their students for their competition rounds.

\section{Conclusions}
\label{sec:conclusions}

The Czech Astronomy Olympiad shows how creativity and being welcoming to everyone can make astronomy education more exciting and get more people interested in science. We hope our experiences, methods and Open Access publications can help other educators as well.

\subsection*{Acknowledgements}
{\small
The authors want to acknowledge the entire CzAO organising team for their continued hard work and dedication to education. A special thank you also belongs to the former president of CzAO, Dr~Jan Ko\v{z}u\v{s}ko, who has led this competition for over 15 years. The presented work was partially supported by the Czech Ministry of Education, Youth and Sports through the project 0005/8/SOU/2023
within the programme MSMT-22244/2022-2.
IWAA 2023 was supported by the project 0047/7/NAD/2023.
CzAO also thanks Planetum and the Silesian University Institute of Physics in Opava for their support of this competition.
}


\end{document}